\documentclass[twocolumn,showpacs,preprintnumbers,amsmath,amssymb,superscriptaddress]{revtex4}
\usepackage{graphicx}% Include Fig. files
\usepackage{dcolumn}% Align table columns on decimal point
\usepackage{bm}% bold math
\usepackage{graphicx}
\usepackage{xspace}
\usepackage{multirow}
\usepackage{natbib}
\usepackage{color}

\renewcommand{\vec}{\mathbf}

\newcommand{\BKFAx}[2]{Ba$_{#1}$K$_{#2}$Fe$_2$As$_2$}

\begin{document}

\preprint{}

%\title{Before and after the oxygen reduction of n-doped cuprates: an ARPES comparison}% Force line breaks with \\
%\title{Electronic origin of the reduction-induced superconductivity in the n-doped high-T$_c$ superconductors}
\title{Observation of an Orbital Selective Electron-Mode Coupling in Fe-Based High-$T_c$ Superconductors}

\author{P. Richard}\email{p.richard@arpes.phys.tohoku.ac.jp}
\affiliation{WPI Research Center, Advanced Institute for Material Research, Tohoku University, Sendai 980-8577, Japan}
\author{T. Sato}
\affiliation{Department of Physics, Tohoku University, Sendai 980-8578, Japan}
\author{K. Nakayama}
\affiliation{Department of Physics, Tohoku University, Sendai 980-8578, Japan}
\author{S. Souma}
\affiliation{WPI Research Center, Advanced Institute for Material Research, Tohoku University, Sendai 980-8577, Japan}
\author{T. Takahashi}
\affiliation{WPI Research Center, Advanced Institute for Material Research, Tohoku University, Sendai 980-8577, Japan}
\affiliation{Department of Physics, Tohoku University, Sendai 980-8578, Japan}
\author{Y.-M. Xu}
\affiliation{Department of Physics, Boston College, Chestnut Hill, MA 02467}
\author{G. F. Chen}
\affiliation{Beijing National Laboratory for Condensed Matter Physics, and Institute of Physics, Chinese Academy of Sciences, Beijing 100080, China}
\author{J. L. Luo}
\affiliation{Beijing National Laboratory for Condensed Matter Physics, and Institute of Physics, Chinese Academy of Sciences, Beijing 100080, China}
\author{N. L. Wang}
\affiliation{Beijing National Laboratory for Condensed Matter Physics, and Institute of Physics, Chinese Academy of Sciences, Beijing 100080, China}
\author{H. Ding}
\affiliation{Beijing National Laboratory for Condensed Matter Physics, and Institute of Physics, Chinese Academy of Sciences, Beijing 100080, China}

\date{\today}% It is always \today, today,

\begin{abstract}
We have performed an angle-resolved photoemission spectroscopy study of the new superconductor \BKFAx{0.6}{0.4}\xspace in the low energy range. We report the observation of an anomaly around 25 meV in the dispersion of superconducting \BKFAx{0.6}{0.4}\xspace samples that nearly vanishes above $T_c$. The energy scale of the related mode (13$\pm$2 meV) and its strong dependence on orbital and temperature indicates that it is unlikely related to phonons. Moreover, the momentum locations of the kink can be connected by the antiferromagnetic wavevector. Our results point towards an unconventional electronic origin of the mode and the superconducting pairing in the Fe-based superconductors, and strongly support the anti-phase $s$-wave pairing symmetry.
\end{abstract}

%\verb+\pacs{#1}+ command.

%Use showkeys class option if keyword
\maketitle

%\section{Introduction}

%Previous angle-resolved photoemission spectroscopy (ARPES) studies revealed an anomaly in the energy dispersion of superconducting cuprates that is believed to play a key role in the pairing mechanism.

The recent discovery of Fe-based superconductors with critical temperatures up to 56 K \cite{Kamihara, Takahashi, Wen, XH_Chen, GF_Chen, Ren} raises the prospect of unconventional superconducting pairing mechanism. While the electronic pairing in conventional superconductors is mediated by lattice vibration modes (phonons), its nature in the Fe-based high-$T_c$ superconductors is unknown. Similarly to copper oxides (cuprates), these materials exhibit anomalously large 2$\Delta/k_BT_c$ ratios \cite{Ding_EPL} and are close to antiferromagnetic instabilities. However, they differ from the copper oxides by the symmetry of the order parameter, which is $s$-wavelike rather than $d$-wave \cite{Ding_EPL}. Whether superconductivity in these materials can be driven by phonons or not is under intense debate and it is of crucial importance to identify signatures of the pairing mechanism itself.  

The many-body interactions responsible for superconductivity are closely related to the electronic energy dispersion and other excitations in the vicinity of the Fermi level ($E_F$). Angle-resolved photoemission spectroscopy (ARPES) is a powerful technique to directly access the electronic dispersion of materials and its interactions with bosonic excitations. Indeed, a direct signature of an electron-mode coupling is an anomaly in the electronic energy dispersion (kink). For example, previous ARPES studies revealed a kink in cuprates \cite{Norman1,Valla1, Kaminski1, Lanzara1}, which is believed to be linked to the pairing. 

We searched anomalies in the low energy scale of the electronic dispersion of \BKFAx{0.6}{0.4}\xspace single-crystals ($T_c$ = 37 K) by performing high-resolution ARPES measurements. In this Letter, we report the observation of a kink around 25 meV. Similarly to the antinodal kink observed in cuprates \cite{Norman1, Kaminski1}, the kink in \BKFAx{0.6}{0.4}\xspace nearly vanishes above $T_c$. In addition to the energy scale of the related energy mode (13$\pm$2 meV), the kink has a strong dependence on orbital and temperature that suggests a non-phononic origin. Furthermore, the observation of the kink only on two Fermi surface (FS) sheets connected by the antiferromagnetic vector $\Vec{Q}_{AF}$ point towards an electronic origin and strongly support the anti-phase $s$-wave pairing symmetry.   

%\section{Experiment}

High-quality single crystals of \BKFAx{0.6}{0.4}\xspace  ($T_c$ = 37 K) have been grown using the flux method \cite{GF_Chen2}. A microwave-driven helium source ($h\nu$ = 21.218 eV and 40.814 eV) and a VG-Scienta 2002 multi-channel analyzer were used to record ARPES spectra with energy and momentum resolutions of 4-7 meV and 0.007-0.01 \AA$^{-1}$, respectively. The samples were cleaved \emph{in-situ} at 15 K and measured with working vacuum better than 5$\times$10$^{-11}$ Torr.  No obvious degradation of the spectra was observed for typical measurements of 3 days.

\begin{figure}[htbp]
\begin{center}
\includegraphics[width=8cm]{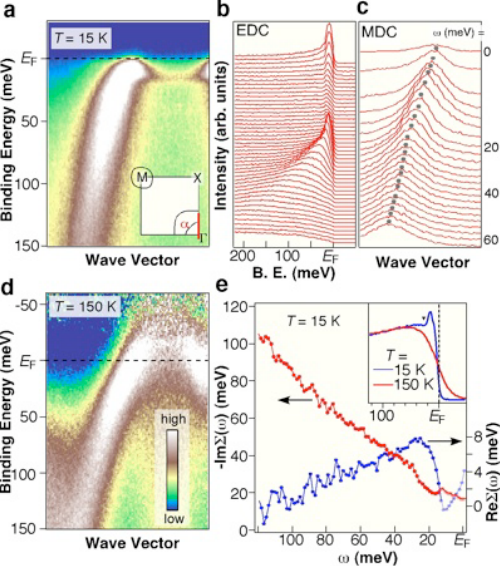}
\caption{\label{fig_alpha}(Color) (a) ARPES intensity plot in the superconducting state (15 K) along a cut crossing the $\alpha$ band. The inset shows the schematic FS of \BKFAx{0.6}{0.4}\xspace with the location of the cut (red). $\Gamma$ (0, 0), X ($\pi$/2, $\pi$/2) and M ($\pi$, 0) are high symmetry points described in the unreconstructed BZ.  (b) Corresponding EDCs.  (c) Corresponding MDCs, in the 0-60 meV binding energy range. Grey dots indicate the maximum position of the peaks. (d) ARPES intensity plot at 150 K divided by a Fermi-Dirac function, recorded along the same cut as in (a). (e) Real and imaginary parts of $\Sigma$($\omega$). Fade colors are used for binding energies smaller than 17 meV since $\Sigma$($\omega$) in this range is complicated by particle-hole mixing due to superconductivity. The inset compares the partial DOS along the cuts measured at 15 K (blue) and 150 K (red), respectively.} 
\end{center}
\end{figure}

%\section{Results and discussion}
In contrast to copper oxides, the iron-based superconductors (ferropnictides) have a complex FS \cite{Ding_EPL,Lu, Liu} with at least three bands crossing $E_F$, called $\alpha$, $\beta$ and $\gamma$, respectively \cite{Ding_EPL}. We first focus on the inner holelike band centered at the $\Gamma$ point, named $\alpha$, which has a 12 meV superconducting gap \cite{Ding_EPL}. Figure \ref{fig_alpha} summarizes the results obtained at 15 K along a cut in momentum space indicated in the inset of Fig. \ref{fig_alpha}(a). The ARPES intensity plot displayed in panel (a) shows a clear discontinuity in the electronic dispersion around 25 meV. This anomaly is emphasized in the corresponding energy distribution curves (EDCs) shown in Fig. \ref{fig_alpha}(b), which represent the photoemission intensity as a function of binding energy at a fixed momentum. The sharpness of the peaks suggests a very good sample quality. We find a dip in the EDCs whose energy is always fixed around 25 meV, characteristic signature of an electron-mode coupling. Figure \ref{fig_alpha}(c) shows the plots of photoemission intensity at 15 K as a function of momentum at a fixed energy, commonly called momentum distribution curves (MDCs). 

In the presence of many-body interactions, an electron-mode coupling is usually described by a complex self-energy $\Sigma$($\omega$). Following a common practice in ARPES \cite{Johnson}, we approximated the real part of $\Sigma$($\omega$) induced by the coupling by the difference between the MDC dispersions obtained at 15 K and 150 K, the latter used as an approximation of the ``bare" band dispersion without electron-mode coupling. The intensity plot of the 150 K data used for the extraction of $\Sigma$($\omega$) is given in Fig. \ref{fig_alpha}(d). The data at 150 K have been normalized between -40 and 150 meV after division by the Fermi-Dirac function in order to minimize the effect of photoemission matrix element suppression at the $\Gamma$ point and facilitate the visualization of the band above $E_F$. Our approach is justified by the absence of kink and energy gap at 150 K. Moreover, the anomaly is observed in the partial density of states (DOS) at 15 K obtained by integrating the EDCs over a wide momentum range along the measured cut, while it is completely absent at 150 K (inset of Fig. \ref{fig_alpha}(e). In addition, the thermal broadening of the Fermi cutoff at 150 K allows us to determine that the $\alpha$ band tops at 20 meV above $E_F$. The imaginary part of $\Sigma$($\omega$) is approximated by the half-width at half-maximum of the Lorenztian functions used to fit the MDCs, multiplied by the velocity of the bare band at each energy $\omega$.  The results of the self-energy analysis are shown in Fig. \ref{fig_alpha}(e). Both the real and imaginary parts of $\Sigma$($\omega$) exhibit a well defined anomaly at a position similar to the dip, as expected in the presence of an electron-mode coupling. Particularly, the drop in the absolute value of the imaginary part of $\Sigma$($\omega$) indicates a reduced quasiparticle scattering rate. In simple models, a coupling with an Einstein mode of energy $\Omega$ in the presence of a superconducting gap $\Delta$ will be observed at a binding energy $\Omega+\Delta$ \cite{Norman2}. Subtracting the 12 meV superconducting gap of the $\alpha$ band \cite{Ding_EPL} from the anomaly energy, we assign to the mode an energy of $\Omega$ = 13 $\pm$ 2 meV.

%\newpage

\begin{figure}[htbp]
\begin{center}
\includegraphics[width=8cm]{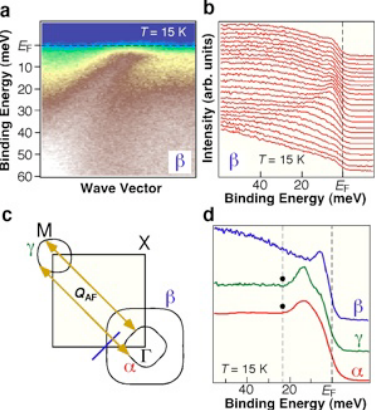}
\caption{\label{orbital_dep}(Color) (a) ARPES intensity plot in the superconducting state (15 K) along a cut crossing the $\beta$ band. (b) EDCs corresponding to the cut shown in panel (a). (c) Schematic Fermi surface of \BKFAx{0.6}{0.4}\xspace with the location of the cuts presented in panel (a). We also traced the AF wavevector $\vec{Q}_{AF}$, which connects the $\alpha$ and $\beta$ bands. (d) Partial DOS associated with the $\alpha$, $\beta$ and $\gamma$ bands. The dots represent the position of the dip.}
\end{center}
\end{figure}

We now address the important question of the orbital dependence of the dispersion anomaly. In order to investigate this issue, we measured ARPES spectra along a cut crossing the $\beta$ FS (outer holelike FS centred at the $\Gamma$ point). The corresponding ARPES intensity plot and EDCs are given in Fig. \ref{orbital_dep}(a) and Fig. \ref{orbital_dep}(b), respectively, and the location of the cut in momentum space is indicated in Fig. \ref{orbital_dep}(c). Surprisingly, the $\beta$ band exhibits a kink neither at 25 meV nor at $\Delta_{\beta}$+$\Omega$ = 19 meV, where $\Delta_{\beta}$ = 6 meV \cite{Ding_EPL}. The partial DOS around the $\beta$ band Fermi wavevector supports this observation, in contrast to the $\alpha$ and $\gamma$ bands, which both show a dip around 25 meV in the partial DOS (Fig. \ref{orbital_dep}(d)). This behavior is not simply explained by a conventional isotropic electron-phonon coupling and suggests the orbital selective nature of the mode. The mode is preferentially coupled to the electrons in the $\alpha$ and $\gamma$ bands, which are well connected via the antiferromagnetic (AF) momentum transfer $\Vec{Q}_{AF}$ = ($\pi$, 0), here expressed in the unreconstructed Brillouin zone (BZ), or ($\pi$, $\pi$) in the reconstructed BZ. We illustrate this scenario in Fig. \ref{orbital_dep}(c) and stress that a less than perfect nesting is compensated, in the presence of inelastic scattering, by the flatness and shallowness of the $\gamma$ band \cite{Ding_EPL}.

%\newpage

\begin{figure}[htbp]
\begin{center}
\includegraphics[width=8cm]{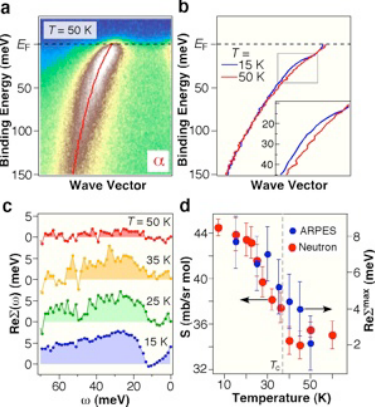}
\caption{\label{temperature_dep}(Color) (a) ARPES intensity plot in the normal state (50 K) along a cut crossing the $\alpha$ band (see inset of Fig. \ref{fig_alpha}(a)). (b) Comparison of the MDC fit dispersion for the $\alpha$ band in the superconducting state (15 K) and in the normal state (50 K) for the cut presented in panel (a). The inset shows a zoom around the kink. (c) Temperature dependence of the real part of the self-energy referred from the MDC fit dispersion at 150 K. (d) Maximum value of the real part of the self-energy (blue) plotted as a function of temperature. The ARPES results are compared to the neutron scattering intensity of the 14 meV spin resonance (red) located at the AF wavevector \cite{Christianson}. The dashed line indicates the critical temperature.}
\end{center}
\end{figure}

To characterize further the kink in the dispersion, we performed a temperature dependence of the ARPES measurements associated with the $\alpha$ band. The results are summarized in Fig. \ref{temperature_dep}. In panel (a), we show the spectra obtained at 50 K. The 50 K data (normal state) differ from the 15 K ones not only by the absence of a superconducting gap, but also by a marked suppression of the kink, suggesting a strong correlation between the electron-mode coupling and the superconducting state. The suppression of the kink is also clearly seen in Fig. \ref{temperature_dep}(b), where we overlap the dispersion extracted from the MDC analysis at 15 K and 50 K. The temperature dependence of Re $\Sigma$($\omega$) is given in Fig. \ref{temperature_dep}(c). The peak in Re $\Sigma$($\omega$) characterizing the kink is strongly suppressed with temperature and nearly vanishes above $T_c$. Similarly, the maximum value of Re $\Sigma$($\omega$), which is given in Fig. \ref{temperature_dep}(d), has a strong temperature dependence and tends to disappear above $T_c$ ($T$ = 50 K). 
  
%\newpage

Calculations suggest that the Fe contribution to the phonon DOS in LaFeAsO$_{1-x}$F$_x$ has a peak centered at 12 meV \cite{Singh,Boeri}, close to the value found experimentally by inelastic neutron scattering on the same compound \cite{Christianson_DOS}. However, besides the unusual orbital dependence of the kink, the phononic nature of the mode would not be able to explain satisfactorily at least two other points: 1) unlike the ARPES kink, the experimental phonon DOS does not show any strong temperature dependence \cite{Christianson_DOS}; 2) the predicted coupling strength $\lambda\approx 0.21$ is too small \cite{Singh,Boeri} to explain the present ARPES results and the optical measurements \cite{Yang}.

A more natural explanation is that the mode has an electronic origin. It has been suggested, especially from the orbital dependence of the superconducting gap \cite{Ding_EPL}, that pairing in the ferropnictides is favored by the nesting of the $\alpha$ and $\gamma$ bands, centered respectively at the $\Gamma$ and M points \cite{Mazin,Kuroki,Wang}. They have almost same gap magnitude, which is twice as much as the gap observed for the $\beta$ band \cite{Ding_EPL}. Interestingly, a recent inelastic neutron scattering study of \BKFAx{0.6}{0.4}\xspace revealed a spin resonance centered around 14 meV at the AF wavevector \cite{Christianson}. As shown in Fig. \ref{temperature_dep}(d), the temperature dependence of the corresponding neutron scattering intensity agrees remarkably well with our result for the maximum value of Re $\Sigma$($\omega$), pointing towards a same origin. Indeed, the opening of a superconducting gap $\Delta$ in the strong coupling regime is naturally accompanied by the opening of an energy gap in the electronic spin excitations, which can give rise to a resonance mode due to divergence in the magnetic spin susceptibility (resonance mode) \cite{Chubukov,Korshunov,Maier}. In the presence of such a feedback process, induced kinks in the dispersion are expected around 2$\Delta$, in agreement with our result for the $\alpha$ and $\gamma$ bands. Within our resolution, no obvious kink is found for the $\beta$ band at 2$\Delta$ (12 meV), suggesting the possible orbital selectivity of the feedback process. We caution that a kink may also be induced by band folding ($\alpha\leftrightarrow\gamma$) resulting from the enhanced AF fluctuations below $T_c$, in which case the kink would be better described in terms of hybridization. In any of these scenarios though, or even in a combination of them, the kink keeps a purely electronic origin. 

At this point it is interesting to compare the kink reported in this letter with the kink observed in the cuprates. While phonons have been proposed to explain this feature \cite{Lanzara1}, some experiments and calculations contradict this hypothesis \cite{Norman1,Johnson,Terashima,Giustino} and the nature of the kink in cuprates remains an open question. Although a one-to-one correspondence between the Fe-based and Cu-based superconductors is not expected, the kinks observed in the two families of compounds behave similarly: both are observed at an energy close to 2$\Delta$ at locations that can be linked by an AF wavevector, in addition to have a strong temperature dependence and disappear above $T_c$. It is therefore tempting to assume that they also share the same origin. More significantly, the emergence of an AF resonance mode has been intimately associated with the opposite pairing phases between the connected FS sections in both cuprates \cite{Chubukov} and Fe-based superconductors \cite{Korshunov,Maier}. Thus, the observation of an orbital-selective mode coupling strongly supports the idea of antiphase $s$-wave pairing in the new high-$T_c$ superconductors.

In summary, we found a kink around 25 meV in the electronic energy dispersion of the new superconductor \BKFAx{0.6}{0.4}\xspace that we assign to a mode of energy 13 $\pm$ 2 meV. The mode nearly vanishes above $T_c$ and is observed only on the $\alpha$ and $\gamma$ FS, which are well connected by the antiferromagnetic wavevector. These observations of a non-ubiquitous kink with strong temperature dependence strongly suggest an unconventional mechanism for the mode of purely electronic origin.  

%\begin{acknowledgments}
We thank T. Arakane for experimental assistance, and X. Dai, Z. Fang and Z. Wang for valuable discussions. This work was supported by grants from JSPS, JST-CREST, MEXT of Japan, the Chinese Academy of Sciences, NSF, Ministry of Science and Technology of China, and NSF of US.
%\end{acknowledgments}

\bibliography{biblio_en}

\end{document}